# Orbital occupation selects structural dimensionality in binary transition-metal oxides

Deping Guo[1,2,3], Renhong Wang[2,3], Cong Wang[2,3], Meng Gao[4], Wu Zhou[4], Yanning Zhang[5], Fei Pang[2,3], Zhihai Cheng[2,3], and Wei Ji[2,3,5*]

[1]*College of Physics and Electronic Engineering, Center for Computational Sciences, Sichuan Normal University, Chengdu, 610101, China*

[2]*Beijing Key Laboratory of Optoelectronic Functional Materials & Micro-Nano Devices, School of Physics, Renmin University of China, Beijing 100872, China*

[3]*Key Laboratory of Quantum State Construction and Manipulation (Ministry of Education), Renmin University of China, Beijing, 100872, China*

[4]*School of Physical Sciences, University of Chinese Academy of Sciences, Beijing 100049, China*

[5]*Institute of Fundamental and Frontier Sciences, University of Electronic Science and Technology of China, Chengdu, 610054, China*

[*]*Corresponding authors.* Email: wji@ruc.edu.cn (W.J.).

**Abstract:** Orbital-lattice coupling in transition-metal oxides is usually discussed within a given bonding framework, where orbital occupation is intertwined with local coordination, strain, or symmetry breaking. Here, we show that orbital occupation can also select the bonding framework itself, thereby determining structural dimensionality. Using first-principles calculations, we identify CrO as a prototype in which the single active 3$d$ $e_g$ electron of high-spin $Cr^{2+}$ gives rise to two competing orbital-structure states. The $d_{x^2-y^2}$ occupation favors a three-dimensionally connected covalent phase, whereas the $d_{z^2}$ occupation stabilizes a weakly coupled layered phase. Constrained-occupation calculations show that increasing the $d_{z^2}$ filling continuously contracts the in-plane lattice while expanding the structure along the layer normal. The two phases exhibit distinct magnetic ground states and ferroelastic responses. Moreover, the layered phase is robust against exchange-correlation functional and on-site ($U$) variations, remains dynamically stable down to the monolayer limit, and has a low exfoliation energy of 46 meV/Å$^2$. Extending the analysis across related 3$d$ binary oxides reveals a filling-dependence relation between accessible orbital filling and the preference for 2D or 3D connected bonding motifs, providing a microscopic basis for exploring low-dimensional oxide materials.

Transition-metal oxides (TMOs) host a remarkable range of crystal structures and electronic phases because their localized *d* electrons are strongly coupled to charge [1], spin [2–4], orbital [5,6], lattice [7], and valley [8] degrees of freedom. Reducing their dimensionality from three-dimensional (3D) bulk to two-dimensional (2D) layers can reshape orbital hybridization [9], reduce electronic bandwidth, enhance electron correlations [10], and modify magnetic exchange interactions [11,12] and ferroic responses [13]. These low-dimensionality induced changes can stabilize structural phases and electronic states that are inaccessible in bulk crystals [14–17]. Moreover, advances in manipulating two-dimensional materials, particularly electrostatic gating, offer direct approaches to tune carrier density and examine filling-dependent electronic and symmetry-breaking phenomena in metal oxides. However, even in the simplest binary TMOs, reaching their low-dimensional limit remains challenging because most known TMOs favor 3D bonding networks rather than 2D layered structures [18–24]. Therefore, existing strategies for obtaining 2D oxides usually rely on external constraints applied to non-layered oxides, such as substrate stabilization in epitaxy, interfacial engineering, or freestanding-membrane approaches [25,26]. This raises a more fundamental question: can structural dimensionality of TMOs be selected internally by an electronic degree of freedom?

Among the electronic degrees of freedom available in TMOs, orbital occupation is closely tied to the directionality of metal-oxygen bonding. In open-shell TMOs, orbital occupation of *d* electrons determines the anisotropic distribution of valence charge and the population of metal-oxygen antibonding states, thereby coupling electronic states to lattice distortions and local structures. Such orbital-lattice coupling underlies Jahn-Teller distortions and orbital ordering, where electronic and structural degrees of freedom evolve cooperatively within a given bonding network. However, these studies have largely focused either on orbital ordering and lattice distortions within an established bonding network, or on tuning orbital occupation through external perturbations, such as strain [27], doping [28], or symmetry breaking [29,30]. Much less explored is whether orbital occupation can instead alter the global bonding network itself, thereby selecting structural dimensionality of oxides. Among all open-shell $d$ electron occupation configurations, transition metal cations with a single $e_g$ electron provide a simple setting to examine this possibility, as the active $e_g$ electron may occupy either the $d_{x^2-y^2}$ [Fig. 1(a)] or $d_{z^2}$ orbital [Fig. 1(b)]. Owing to their distinct directional characters, these occupations can weaken metal-oxygen bonds within the basal plane [Fig. 1(c)] and along the layer normal [Fig. 1(d)], respectively, leading to lattice expansion in the two distinct directions.

Here, using first-principles calculations, we demonstrate this possibility in CrO, where high-spin $Cr^{2+}$ has a single active $3d$-$e_g$ electron. We find that two $e_g$ occupations give rise to two distinct bonding geometries. The occupation of $d_{x^2-y^2}$ is associated with an in-plane-expanded, rock-salt-like 3D phase [31] and the occupation of $d_{z^2}$ stabilizes an out-of-plane-expanded square-lattice 2D layered phase. The latter exhibits van der Waals-like (vdW-like) characteristics, including a large interlayer spacing, low exfoliation energy, weak layer dependence of its electronic structure, and dynamical stability down to the monolayer limit. A recently experiment realized freestanding monolayer and bilayer CrO, including bilayers with multiple rotational registries [32], consistent with the monolayer stability and weakly coupled layered character found here. Our calculations also provide a microscopic explanation for why CrO can support such a layered bonding motif rather than an extended 3D Cr-O network. The 3D and 2D phases further exhibit different magnetic ground states and ferroelastic responses, revealing a pronounced coupling among orbital occupation, lattice geometry, and magnetism. Extending this analysis to related $3d$ binary TMOs shows that the competition between 3D and 2D phases follows the accessible $e_g$ orbital filling configurations.

Our density functional theory (DFT) calculations were carried out using the generalized gradient approximation for the exchange-correlation potential [33], the projector augmented wave method [34] and a plane-wave basis set as implemented in the Vienna ab-initio simulation package (VASP) [35,36]. Dispersion corrections were made at the van der Waals density functional (vdW-DF) level [37] with the optB86b functional for the exchange potential (optB86b-vdW) [38] in all structural relaxations. On-site Coulomb interactions for the *d* orbitals of Cr, Mn, Fe, Co, Ni, and Cu were considered using a DFT+U method [39]. An effective *U* value of 3.0 eV was adopted for all elements, comparable to those used in the literature [40,41].

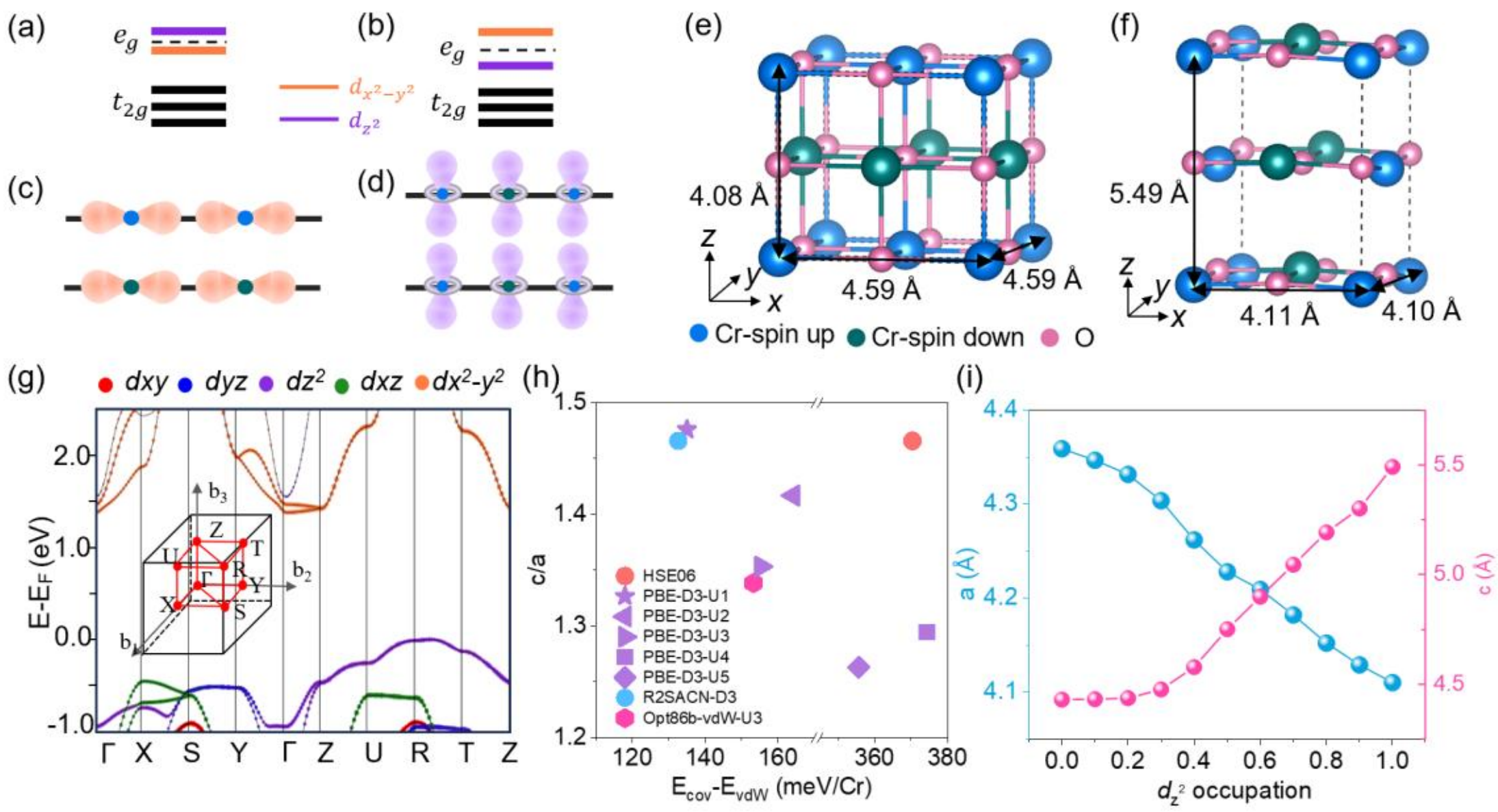

FIG. 1. Schematic illustrations of orbital occupations A (a) and B (b). The black dashed line represents the Fermi energy level. Illustration of orbital occupations A (c) and B (d) in real-space. Blue and green spheres represent Cr atoms with spin-up and spin-down orientations, respectively. The atomic structures and magnetic configurations of covalent- (e) and vdW-like CrO (f). (g) Orbital-projected band structures of vdW-like CrO, with the high-symmetry paths sampling the Brillouin zone (inset). (h) Robustness of the v-CrO phase under different exchange-correlation functionals and on-site *U* values, showing the relative energy of v-CrO with respect to c-CrO and the corresponding out-of-plane to in-plane lattice ratio *c*/*a*. (i) Lattice constants *a* (blue) and *c* (pink) as a function of the $d_{z^2}$ orbital occupancy.

Figure 1(e) shows the fully relaxed bulk CrO atomic structure associated with preferentially occupied in-plane $d_{x^2-y^2}$ orbital (orbital occupation A, OO-A), as confirmed by the orbital-projected electronic band structures (Fig. S1). Under OO-A, bulk CrO adopts a rocksalt-like structure with a tetragonal distortion and metallic bandstructure (Fig. S1), close to a high-temperature cubic phase reported experimentally [31]. The optimized lattice constants are $a = b = 4.59$ Å, and $c = 4.08$ Å, corresponding to Cr-O bond lengths of 2.30 (in-plane) and 2.04 (out-of-plane) Å, both consistent with typical covalent Cr-O bonding [42]. We thus denote this structure as covalent CrO (c-CrO).

Replacing the occupation of the $d_{x^2-y^2}$ orbital with the $d_{z^2}$ orbital (OO-B) relaxes bulk CrO into a markedly different layered structure [Fig. 1(f)] with a band gap of 1.37 eV, as confirmed by the orbital-projected band structure [Fig. 1(g)]. In this phase, the crystal expands strongly along the out-of-plane direction, with optimized lattice constants $a$ = 4.11 Å, $b$ = 4.10 Å and $c$ = 5.49 Å. The resulting out-of-plane Cr-O distance is 2.75 Å, 1.38 times the sum of the covalent radii of Cr and O, indicating a much weaker interlayer interaction than that in the covalent phase. We therefore denote this structure as vdW-like CrO (v-CrO). The energetic preference for v-CrO over c-CrO is robust across the tested exchange-correlation functionals and on-site $U$ values from 1.0 to 5.0 eV. In all cases, the relaxed v-CrO phase retains the $d_{z^2}$ orbital occupation [Figs. 1(h) and S2]. In addition, the Cr sublattice in each CrO layer is buckled by 0.28 Å, further distinguishing the vdW-like v-CrO from covalent c-CrO phase. To directly test the orbital-structure correlation, we performed constrained-occupation structural relaxations in which the occupation ratio of $d_{z^2}$ orbitals was kept fixed in each relaxation. Continuously varying the ratio from 0 to 1 (increasing the $d_{z^2}$ portion) decreases $a$ and increases $c$, driving the structure toward v-CrO. The monotonic structural response with $d_{z^2}$ filling supports orbital-filling-driven dimensionality selection.

The contrast between c-CrO and v-CrO extends beyond bonding geometry to their preferred magnetic ground orders. The c-CrO favors an A-type antiferromagnetic [AFM1, as illustrated in Fig. 1(e)] configuration among all three configurations considered, which is at least 30.0 meV/Cr lower in energy than the other two (Fig. S3). Bulk v-CrO, however, prefers an in-plane Néel AFM state (AFM2), as shown in Fig. 1(f), which is more than 82.8 meV/Cr lower in energy than the other two competing magnetic states (Fig. S4). Notably, the magnetic order is tightly coupled to the structural anisotropy in both c-CrO and v-CrO. Specifically, the contracted lattice vector is always oriented normal to the ferromagnetic plane in the c-CrO phase [Fig. 1(e)], while the expanded lattice vector is always perpendicular to the Néel antiferromagnetic plane in the v-CrO phase [Fig. 1(f)].

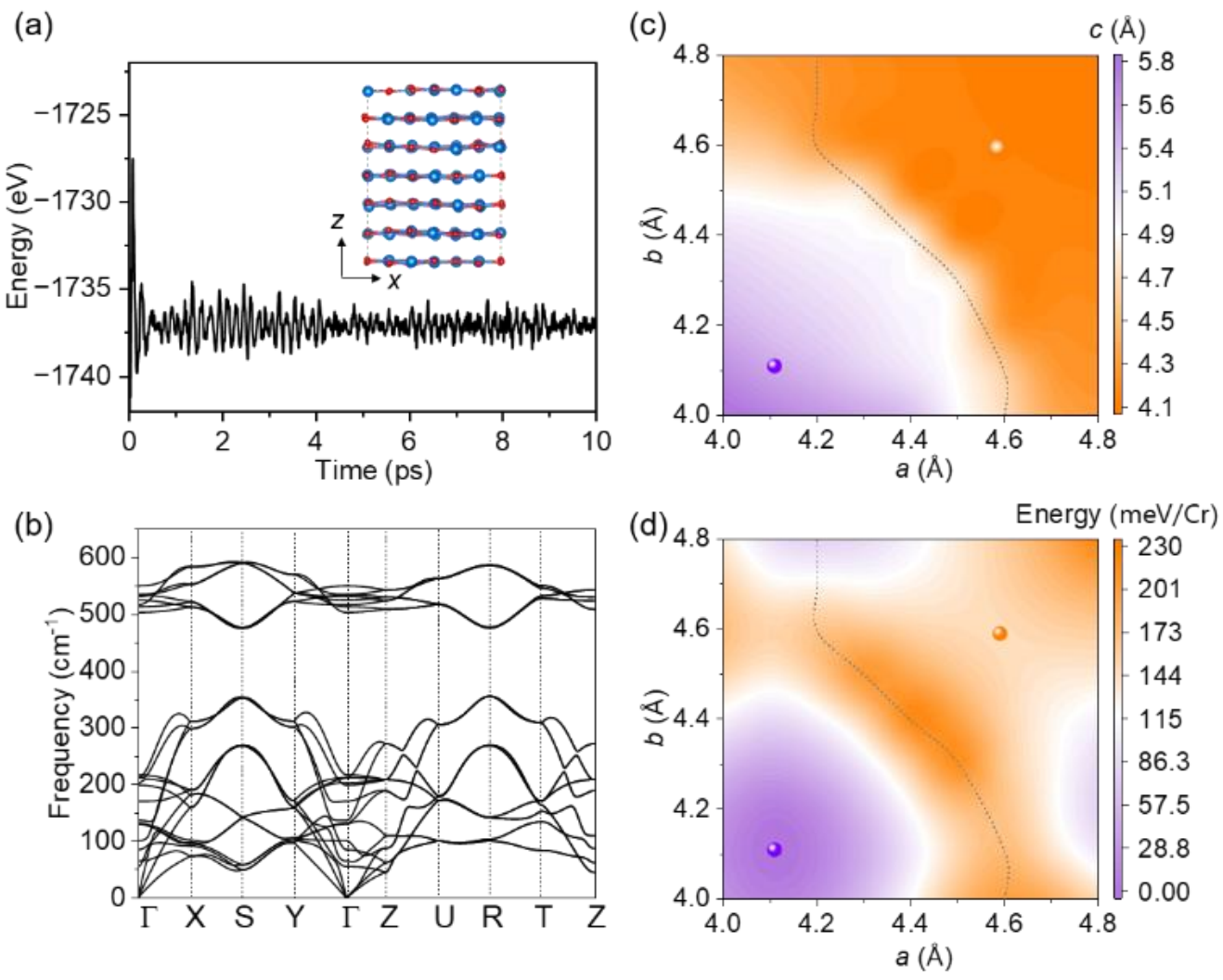

FIG. 2. (a) Temporal evolution of the energy at 300 K, with a structural snapshot (inset) and (b) phonon spectra for the vdW-like CrO. (c) Color map of the out-of-plane lattice constant versus in-plane lattice parameters in bulk CrO. Orange and purple dots correspond to the c-CrO and v-CrO phases, respectively, while the gray dashed line marks the phase boundary between the AFM1 and AFM2 magnetic ground states. (d) Corresponding total energy map as a function of in-plane lattice parameters in bulk CrO.

Because bulk v-CrO structure has not been reported previously, we examined its dynamical and thermal stability using *ab-initio* molecular dynamics simulations, in which the CrO layers remain stable at 300 K for 10 ps [Fig. 2(a)]. Its dynamical stability is also supported by the calculated phonon band structures [Fig. 2(b)], which exhibit no appreciable imaginary phonon frequencies across the Brillouin zone. Energetically, bulk v-CrO is 155 meV/Cr lower than bulk c-CrO (Fig. S5). This better stabilization is accompanied by larger cell volume (92.51 Å$^3$ versus 85.96 Å$^3$) and substantially smaller in-plane area (16.85 Å$^2$ versus 21.07 Å$^2$), consistent with the in-plane and out-of-plane expanded c- and v-CrO phases, respectively. The relative stability of these two phases can also be tuned by the in-plane lattice constant: in-plane contraction favors v-CrO [bottom left region in Fig. 2(c)], whereas in-plane expansion favors c-CrO (top right region). A substantial barrier of 230 meV/Cr separates the optimized v-CrO [purple dot in Fig. 2(d)] and c-CrO (orange dot). Within v-CrO, reorienting the expanded lattice from the $z$ axis to the $y$ axis yields a ferroelastic switching barrier of 76.7 meV/atom (Fig. S6) , which is comparable to the reported values of 63.3 meV/atom for $WTe_2$ [43] and 72 meV/atom for $CoSe_2$ [44].

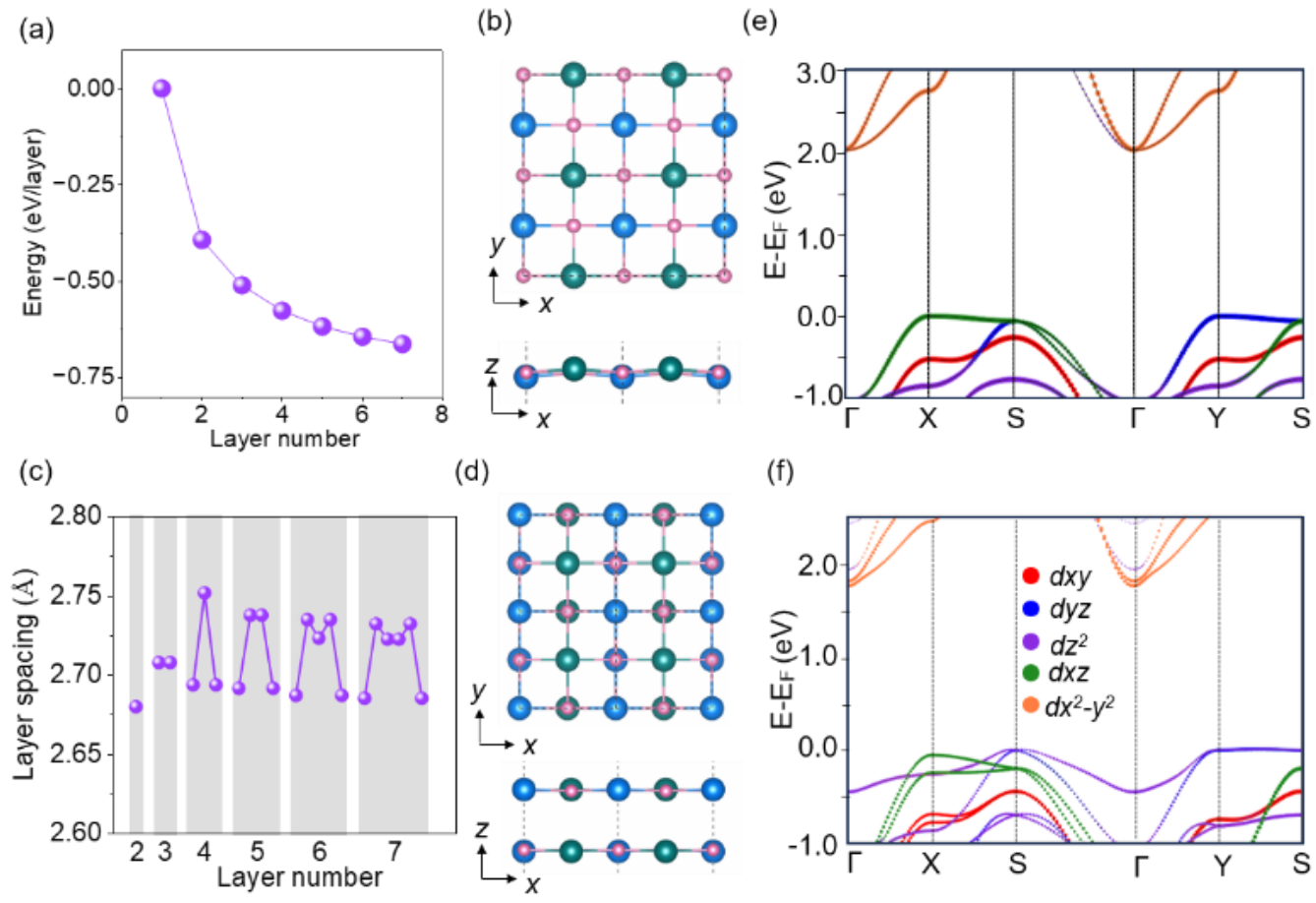

FIG. 3. (a) Plot of the energy variation of few-layer v-CrO as a function of the number of layers. Top (upper panel) and side (lower panel) views of the atomic structures for (b) monolayer and (d) bilayer v-CrO. Blue spheres represent spin-up Cr atoms, green spheres represent spin-down Cr atoms, and pink spheres represent O atoms. (c) Distribution of interlayer spacings in few-layer v-CrO. Corresponding orbital-projected band structures for (e) monolayer and (f) bilayer v-CrO.

We next examine how the layered v-CrO phase evolves with thickness. The layered structural motif remains intact down to the monolayer limit, with progressively reduced energetic stabilization [Fig. 3(a)]. The monolayer is 0.39 eV/Cr higher in energy than bulk v-CrO, yielding an exfoliation energy of 46 meV/Å$^2$, comparable to those of typical 2D materials, such as 38 meV/Å$^2$ for black phosphorus [45]. Meanwhile, the buckling of the Cr plane becomes more pronounced, reaching 0.34 Å in the monolayer [Fig. 3(b)]. The interlayer spacing remains nearly constant in the range of 2.72~2.75 Å with a slightly shorter surface-layer separation of 2.68~2.69 Å [Fig. 3(c)]. Meanwhile, the in-plane lattice constant decreases gradually from 4.11 Å in bulk to 4.04 Å in the monolayer (Fig. S7). This thickness-dependent structural relaxation is accompanied by enhanced quantum confinement, which progressively widens the band gap and yields a maximum value of 2.00 eV in the monolayer limit (Fig. S7).

The v-CrO phase remains dynamically stable from few layers down to the monolayer limit (Fig. S8). The in-plane Néel AFM (N-AFM) configuration remains energetically favored down to the monolayer limit [Fig. 3(b) and Fig. S9]. For the bilayer, the AB stacked configuration, where the Cr atoms in the upper layer are positioned above the O atoms in the lower layer, is more stable than the AA stacked one by over 18 meV/Cr. Although the in-plane N-AFM order is retained, the interlayer magnetic coupling becomes anisotropic: the two layers are FM coupled along the $y$ direction but AFM coupled along $x$ [Fig. 3(d) and Fig. S10]. This anisotropy of interlayer magnetic coupling differentiates the two in-plane lattice constants slightly ($a$=4.07 Å, b=4.08 Å), which are locked to each other.

The orbital-projected band structures of monolayer [Fig. 3(e)] and bilayer v-CrO [Fig. 3(f)] verify the preserved OO-B configuration down to the monolayer limit, exhibiting occupied Cr $d_{z^2}$ [violet in Figs. 3(e) and 3(f)] the unoccupied $d_{x^2-y^2}$ (orange) states. In accordance with the locked magnetic and structural anisotropy, 2L v-CrO also exhibits a pronounced in-plane electronic anisotropy in its valence-band-edge electronic structures between the Γ-X and Γ-Y paths [Fig. 3(f)]. The same anisotropy is retained in thicker few-layer structures (Fig. S11).

The orbital-filling rule identified in CrO extends to other 3$d$ binary oxide bilayers, as summarized in Fig. 4(a) and S12. For other 3$d$ TMOs ranging from CrO to CuO, the preference for a 2D v- or 3D c-phase tracks the occupation of the $e_g$ manifold. Bilayer MnO strongly favors the half-filled high-spin $d^5$ configuration, which suppresses any analogous $d_{z^2}$-driven tendency toward layer separation and instead favors a covalent phase only. In bilayer FeO, CoO, and NiO, the stability of the layered phase is governed by whether the $e_g$ manifold can sustain a simple directional occupation. FeO supports a singly occupied minority-spin $d_{z^2}$ state in the v-phase, which favors layer separation. In CoO and NiO, however, the additional $d$ electrons force the v-phase into mixed $d_{z^2}$-$t_{2g}$ occupations, introducing competing crystal-field and lattice-distortion energies that destabilize the layered geometry. Consequently, the 3D c-phases with $t_{2g}$-dominated occupations become energetically favored. However, CuO represents an intermediate case in this filling trend, as it retains a dominant $d_{z^2}$ occupation with partial $d_{x^2-y^2}$ filling, allowing it to preserve the layered tendency observed in CrO and FeO [Fig. 4(a) and Figs. S13-S14]. The correlation between orbital occupation and the selected structural dimensionality remains robust across different $U$ values (Fig. S15).

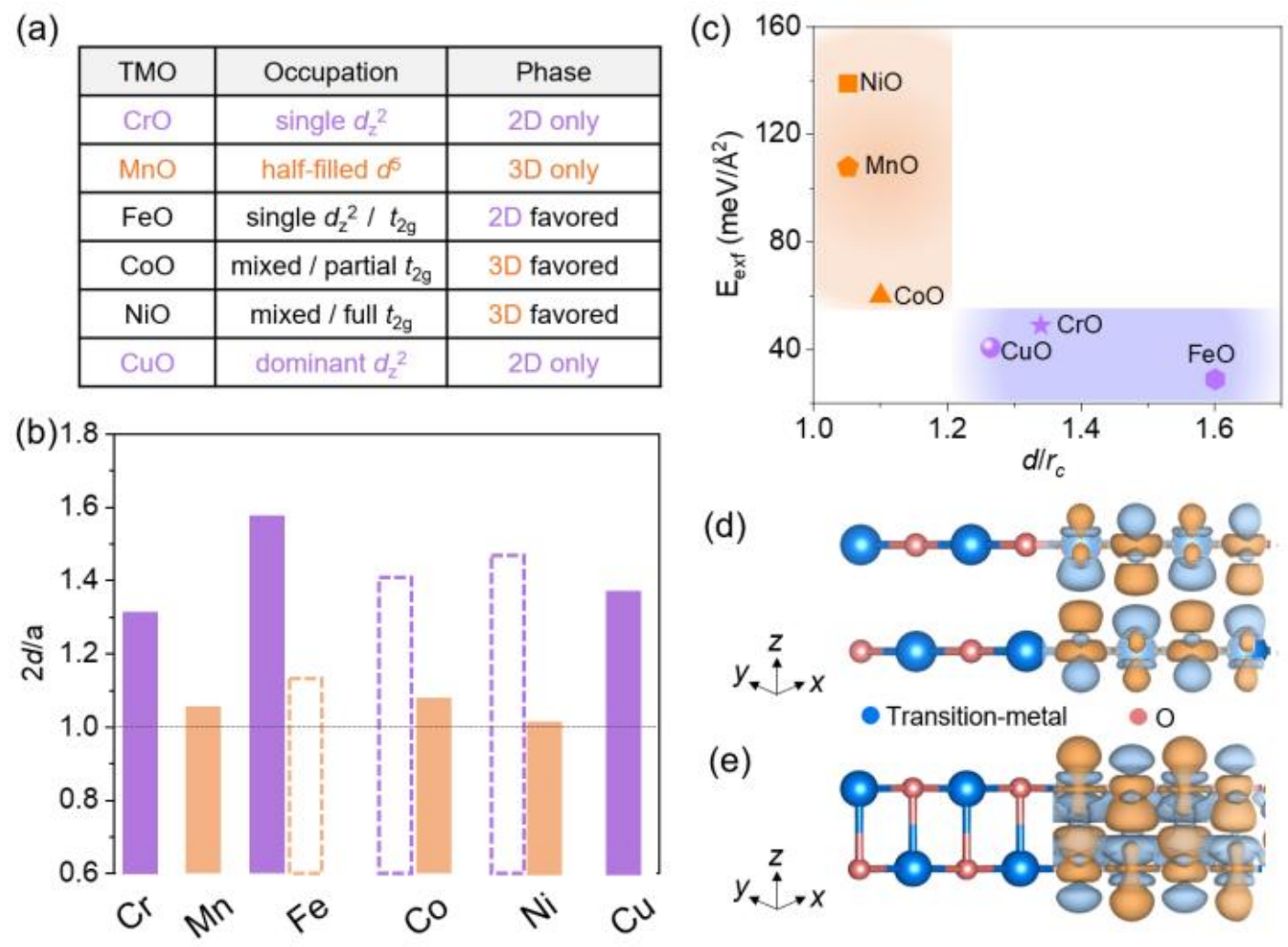

| TMO | Occupation | Phase |
| --- | --- | --- |
| CrO | single $d_z^2$ | 2D only |
| MnO | half-filled $d^5$ | 3D only |
| FeO | single $d_z^2$ / $t_{2g}$ | 2D favored |
| CoO | mixed / partial $t_{2g}$ | 3D favored |
| NiO | mixed / full $t_{2g}$ | 3D favored |
| CuO | dominant $d_z^2$ | 2D only |

FIG. 4. (a) Table of considered binary TMOs, likely occupation configurations of their $d$ orbitals, and favored structural phases. (b) Ratios of twice the interlayer distance ($2d$) to lattice constant $a$ for these oxides. Purple and orange bars represent the structures in the v- and c-phases, respectively. Dashed bars indicate metastable configurations. (c) Exfoliation energy as a function of the normalized interlayer distance $d/r_c$, where $r_c$ is the sum of the covalent radii of interfacial atoms. Purple and orange symbols correspond to v- and c-TMO phases, respectively. Differential charge density (DCD) for (d) v-CuO and (e) c-MnO bilayers. Orange and blue isosurface contours represent charge accumulation and depletion, respectively. The isosurface value is $8\times10^{-4}$ $e$/bohr$^3$.

Figure 4(b) summarizes the normalized out-of-plane to in-plane ratio ($2d/a$) for the nine obtained TMO bilayers. Bilayers FeO, CoO, and NiO support both v- and c-phases as locally accessible structures, indicating a competition between the two bonding motifs. Bilayer CoO provides a representative example of this tunable competition: compressive in-plane and out-of-plane strains selectively stabilize the v- and c-phases, respectively (Fig. S16). Figure 4(c) further connects the orbital-filling trend with interlayer bonding characteristics by plotting the exfoliation energies of the energetically favored oxide phases against the normalized interlayer distance. Bilayer CrO, CuO, and FeO (purple symbols) exhibit large normalized interlayer distances ($d/r_c > 1.26$) and low exfoliation energies of approximately 40 meV/Å$^2$ [Fig. 4(c)], consistent with their (nearly) single $d_z^2$ occupation and vdW-like weak interlayer coupling. In contrast, bilayer MnO, CoO, and NiO (orange symbols) show smaller interlayer separation ($d/r_c$< 1.11) and substantially larger exfoliation energies, indicating stronger interlayer hybridization and more covalent character.

The real-space distinction between these two bonding regimes is further illustrated by the differential charge densities of 2L v-CuO [Fig. 4(d)] and 2L c-MnO [Fig. 4(e)]. In 2L v-CuO, charge redistribution is largely confined within the individual layers, with minimal interlayer charge variation, consistent with a vdW-like layered structure. In 2L c-MnO, by contrast, charge redistribution extends continuously across the interlayer region, indicating appreciable interlayer hybridization and chemical bond formation. Together, these observations indicate that orbital filling determines whether binary oxide bilayers favor weakly coupled layered or strongly connected three-dimensional bonding motifs.

In summary, our first-principles calculations establish orbital occupation as an internal variable for selecting bonding topology and structural dimensionality in binary transition-metal oxides. In CrO, the single active $e_g$ electron of high-spin $Cr^{2+}$ gives rise to two coupled orbital-structure states: $d_{x^2-y^2}$ occupation favors a 3D connected covalent c-phase, whereas $d_{z^2}$ occupation stabilizes a 2D layered v-phase with vdW-like characteristics. The energetic preference, orbital occupation, and magnetic ground state of the layered v-phase remain robust against the tested exchange-correlation functionals and on-site $U$ values. Constrained-occupation structural relaxations further directly link increasing $d_{z^2}$ filling to expansion along the layer normal. The layered v-CrO phase remains dynamically stable down to the monolayer limit and shows a low exfoliation energy and weak thickness dependence of its structural and electronic properties.

Extending the analysis to related 3$d$ binary oxide bilayers reveals that the accessibility and stability of 2D layered and 3D covalent phases follow the accessible $e_g$ filling configurations determined by electron count, identifying FeO and CuO as additional vdW-like layered oxide candidates. Our results place orbital occupation as a central variable in determining bonding topology, extending orbital-lattice physics from local coordination distortions to the dimensionality of oxide frameworks.

## Acknowledgements

We gratefully acknowledge the financial support from the National Natural Science Foundation of China (Grants No. 92477205 and No. 52461160327), the National Key R&D Program of China (Grant No. 2023YFA1406500) and the Sichuan Science and Technology Program (Grant No. 2026NSFSC0767). Calculations were performed at the Physics Lab of High-Performance Computing (PLHPC) and the Public Computing Cloud (PCC) of Renmin University of China, and Hefei Advanced Computing Center.